\documentclass{ws-igtr}

\usepackage{amssymb}
\usepackage{amsfonts}
\usepackage{latexsym}
\usepackage{bbold}
\usepackage[mathscr]{euscript}
\usepackage[cmex10]{amsmath}
\usepackage{graphicx}
\usepackage{algorithm}
\usepackage{algorithmic}
\usepackage{multirow}
\usepackage{caption}
\usepackage{color} 
\newtheorem{cexample}{Counter Example}[section]

\begin{document}


\catchline{}{}{}{}{}

\title{ ON THE NASH STABILITY IN HEDONIC COALITION FORMATION GAMES }

\author{CENGIS HASAN }

\address{School of Informatics, The University of Edinburgh\\ 
10 Crichton StreetEdinburgh, EH8 9AB, UK\\ 
\email{chasan@inf.ed.ac.uk} }

\author{JEAN-MARIE GORCE}

\address{University of Lyon, INSA-Lyon, 6 Avenue des Arts 69621 Villeurbanne Cedex, France\\
\email{jean-marie.gorce@insa-lyon.fr} }

\author{EITAN ALTMAN}

\address{Inria, 2004 Route des Lucioles, 06902 Sophia-Antipolis Cedex, France\\
\email{eitan.altman@inria.fr}}

\maketitle


\begin{abstract}
This paper studies \textit{the Nash stability} in hedonic coalition 
formation games. We address the following issue: for a general 
problem formulation, is there any utility allocation method ensuring 
a Nash-stable partition? We propose the definition of \textit{the 
Nash-stable core}. We study the conditions for having a 
non-empty Nash-stable core. More precisely, we show how relaxed 
efficiency in utility sharing method allows to ensure a non-empty 
Nash-stable core.
\end{abstract}


\ccode{}

\section{Introduction}
Any cooperation among agents (players) being able to make strategic 
decisions becomes a \textit{coalition formation game} when the 
players --for various individual reasons-- may wish to belong to a 
relative 
\textit{small coalition} rather than the ``grand 
coalition''\footnote{The grand coalition is the set of all agents.}. 
Players' moves from one to another coalition are governed by a 
set of rules. Basically, a user will move to a new coalition when it 
may obtain a better utility from this coalition. We shall not 
consider in this paper some permission requirements, which means that 
a player is always accepted by a coalition to which the player is 
willing to join. Based on those rules, the crucial question in the 
game context is how a stable partition occurs. In the following, we 
study the \textit{hedonic} coalition formation games and the 
stability notion that we analyze is the \textit{Nash stability}.  

A coalition formation game is called to be \textit{hedonic} if each player's preferences over partitions of players
depend only on the members of his/her coalition. Finding a stable coalition partition is the main question in a coalition formation game. We refer to \cite{bib:Existenceofstabilityinhedoniccoalitionformationgames} discussing the stability concepts associated to \textit{hedonic 
	conditions}. In the sequel, we concentrate on the Nash 
stability. The definition of the Nash stability is quite simple:
\textit{a partition of players is Nash stable whenever no 
	player deviates from his/her coalition to another coalition 
	in the partition}.

%

\subsection{Related Work}
In \cite{bib:AGenericApproachtoCoalitionFormation}, the authors 
propose an approach to coalition formation that focuses on 
simple merge and split stability rules transforming partitions of a 
group of players. The results are parametrized by a preference 
relation between partitions from the point of view of each player.
In 
\cite{bib:AnalgorithmforgeneratingNashstablecoalitionstructuresinhedonicgames},
the problem of generating Nash-stable solutions in coalitional games 
is considered. In particular, the authors proposed an algorithm for 
constructing the set of all Nash-stable coalition structures from 
players' preferences in a given additively separable hedonic game. 
In \cite{bib:Noncooperativeformationofcoalitionsinhedonicgames}, a bargaining procedure of coalition formation in the class of hedonic games, where players' preferences depend solely on the coalition they belong to is studied. The authors provide an example of nonexistence of a pure strategy stationary perfect equilibrium, and a necessary and sufficient condition for existence. They show that when the game is totally stable (the game and all its restrictions have a nonempty core), there always exists a no-delay equilibrium generating core outcomes. Other equilibria exhibiting delay or resulting in unstable outcomes can also exist. If the core of the hedonic game and its restrictions always consist of a single point, it is shown that the bargaining game admits a unique stationary perfect equilibrium, resulting in the immediate formation of the core coalition structure.

In \cite{drezegreenberg}, Dr\`{e}ze and Greenberg introduced the 
hedonic feature in players' preferences in a context concerning local 
public goods. Moreover, purely hedonic games and stability of hedonic 
coalition partitions were studied by Bogomolnaia and Jackson in 
\cite{bogomonlaia}. In this paper, it is proved that if players' 
preferences are \textit{additively separable and symmetric}, then a Nash 
stable coalition partition exists. For further discussion on 
additively separable and symmetric preferences, we refer the reader 
to \cite{SeparablePreferences}. \textit{Top responsiveness}, which is a condition on players' preferences, captures the idea of how each player believes that others could complement her in the formation of research teams. In \cite{bib:Alcalde}, it is shown that top responsiveness is sufficient for the existence of core stable partitions. The work in \cite{bib:Dimitrov} discusses the existence of Nash-stable partitions in hedonic games possessing top responsiveness property. As shown by the authors, imposing a mutuality condition is sufficient for the existence.

In this work, we study the following problem: having coalitions associated with their utilities, which is called as transferable utility games, we seek the answer of how must the coalition utilities be allocated to the players in order to obtain a stable coalition partition. The fundamental question is to determine which 
utility allocation methods may ensure a Nash-stable partition. We first propose the definition 
of \textit{the Nash-stable core} which is the set of all possible 
utility allocation methods resulting in Nash-stable partitions. We 
show that efficient utility allocations where the utility of a group 
is completely shared between his/her members, may have no Nash-stable 
partitions with only a few exceptions. Rather, we prove that 
relaxing the efficiency condition may ensure the non-emptiness of the 
core. Indeed, we prove that if the sum of players' gains within a 
coalition is allowed to be less than the utility of this coalition, 
then a Nash-stable partition always exists.


\section{Hedonic Coalition Formation}
\subsection{Definition}
A coalition formation game is given by a pair $\langle N,\succ\rangle$, where $ N $ is the set of $ n $ \textit{players} and  $\succ = (\succeq_1,\succeq_2, \ldots ,\succeq_n)$ denotes the \textit{preference profile}, specifying for each player $i\in N$ his preference relation $\succeq_i$, i.e. a reflexive, complete and transitive binary relation.

\begin{definition}
	A coalition structure or a \emph{coalition partition} is a  set $\Pi = \{S_1,\ldots,S_l\}$ which partitions the players set $N$, i.e., $\forall k$, $S_k\subset N$ are disjoint coalitions such that $\bigcup_{k=1}^l S_k = N$. Given $\Pi$ and $i$, let $S_{\Pi}(i)$ denote the set $S_k \in \Pi$ such that $i\in S_k$.
\end{definition}

In its partition form, a coalition formation game is defined on the 
set $N$ by associating a utility value $u(S|\Pi)$ to each subset of 
any partition $\Pi$ of $N$. In its characteristic form, the utility 
value of a set is independent of the other coalitions, and therefore, 
$u(S|\Pi)=u(S)$. The games of this form are more restrictive but 
present interesting properties to reach an equilibrium. Practically 
speaking, this assumption means that the gain of a group is 
independent of the other players outside the group. 

Hedonic coalition formation games fall into this category with an additional assumption:
\begin{definition}
	\label{def:hedonic}
	A coalition formation game is said \textit{hedonic} if 
	\begin{itemize}
		\item \emph{the gain of any player depends solely on the members of the coalition to which the player belongs}, and
		\item \emph{the coalitions arise as a result of the preferences of 
			the players over their possible coalitions' set}.
	\end{itemize}
\end{definition}

\subsection{Preference relation}
The preference relation of a player can be defined over a \emph{preference function}. Let us denote by $\pi_i:2^N \rightarrow \Re$ the preference function of player $i$. Thus, player $i$ prefers the coalition $S$ to $T$ iff,
\begin{equation}
\pi_i(S) \geq \pi_i(T) \Leftrightarrow S \succeq_i T.
\end{equation}

We consider the case where the preference relation is chosen to be the utility allocated to the player in a coalition, then $\pi_i(S) = \phi_i^S$ where $\phi_i^S$ refers to the utility received by player $i$ in coalition $S$. 

In the case of transferable utility games (TU games), as we are 
considering in this paper, the utility of a group can be transfered 
among users in any way. Thus, an utility allocation is said 
relatively efficient if for each coalition $S$, the sum of individual 
utilities is equal to the coalition utility: $\sum_{i \in S} \phi_i^S 
= u(S), \forall S\subseteq N$.

Now, if the preferences of a player are \emph{additively separable}, the preference can be even stated with a function characterizing how a player prefers another player in each coalition. This means that the player's preference for a coalition is based on individual preferences. 
This can be formalized as follows:
\begin{definition}
	The preferences of a player are said to be \textbf{additively 
		separable} if there exists a function $v_i:N \rightarrow \Re$ s.t. 
	$\forall S,T\subseteq N$
	\begin{equation}\label{eq:additivelyseparable}
	\sum_{j\in S} v_i(j) \geq \sum_{j\in T} v_i(j) \Leftrightarrow S \succeq_i T
	\end{equation}
	where $v_i(i)$ is normalized and set to $v_i(i)=0$ .
\end{definition}

A profile of additively separable preferences, represented by $(v_i,\ldots,v_n)$, satisfies \emph{symmetry} if $v_i(j) = v_j(i), \forall i,j$.

\subsection{The Nash stability}
The question we address in this paper concerns the stability of this 
kind of games. The stability concept for a coalition formation game 
may receive different definitions. In the literature, a game is 
either said \textit{individually stable, Nash stable, core stable, 
	strict core stable, Pareto optimal, strong Nash stable, strict strong 
	Nash stable}. We refer to 
\cite{bib:Existenceofstabilityinhedoniccoalitionformationgames} for a 
thorough definition of these different stability concepts.

In this paper, we concentrate only on the Nash stability. The definition of the Nash stability for an \textit{hedonic coalition formation game} is simply
\begin{definition}[Nash Stability]
	\label{def:nashstability}
	A partition of players is Nash-stable whenever no player is incentive 
	to unilaterally change his or her coalition to another coalition in 
	the partition which can be mathematically formulated as:
	\begin{quote}
		\it
		partition $\Pi$ is said to be Nash-stable if no player can benefit 
		from moving from his coalition $S_{\Pi}(i)$ to another existing 
		coalition $S_k$, i.e.:
		\begin{equation}
		\forall i,k: S_{\Pi}(i) \succeq_i S_k \cup \{i\};  S_k \in \Pi \cup 
		\{\emptyset\}.
		\end{equation}
	\end{quote}
\end{definition}
Nash-stable partitions are immune to individual movements even when a player who wants to change does not need permission to join or leave an existing coalition \cite{bogomonlaia}.

\begin{remark}
	In the literature (\cite{bogomonlaia,hajdukova}), the stability 
	concepts being immune to individual deviation are \emph{Nash 
		stability, individual stability, contractual individual stability}. 
	Nash stability is the strongest within above. The notion of 
	\emph{core stability} has been used already in some models where 
	immunity to coalition deviation is required \cite{hajdukova}. But the 
	Nash-stable core has not been defined yet at the best of our 
	knowledge. This is what we derive in the next section.
\end{remark}

\begin{remark}[Impossibility of a stability concept]
	In \cite{bib:Barber}, the authors propose some set of axioms which are \textit{non-emptiness, symmetry pareto optimality, self-consistency}; and they analyze the existence of any stability concept that can satisfy these axioms. It is proven that for any game $ |N|>2 $, there does not exist any solution which satisfies these axioms.
	
	In this work, we do not seek finding conditions which satisfy the above-mentioned axioms, but rather, we show how only non-emptiness can be guaranteed using Nash stability as a solution concept.
\end{remark}

\section{The Nash-stable Core}
\subsection{Definition}
Consider a hedonic TU game noted $\langle N,u,\succ\rangle$ (since $u$ is transferable to the players, we consider hedonic coalition formation games based on transferable utility). For the sake of simplicity, the \textit{preference function} $\pi_i$ of player $i$ is assumed to be the gain obtained in the corresponding coalition, i.e., $\pi_i(S) \equiv \phi_i^S, \forall i\in S, \forall S \subseteq N$. We denote the \textit{allocation method} as $\Phi=\{\phi_i^S; \forall i\in S, \forall S \subseteq N\}\in \mathbb{R}^{\kappa}$ where $ \kappa = n2^{n-1} $. Note that any allocation method shall create a preference profile.

Define the mapping $\mathsf{\mathsf{M}} : \mathbb{R}^{\kappa} \rightarrow {\cal P} \cup \emptyset$, where ${\cal P}$ is the set of all possible 
partitions. Clearly, for any preference function, the mapping 
$\mathsf{M}$ finds the set of Nash-stable partition $\Pi$, i.e. 
$\mathsf{M}(\Phi) := \{ \Pi \left| \mbox{Nash-stable} \right.\}$. If 
a Nash-stable partition cannot be found, $ \mathsf{M} $  maps to empty 
set. Moreover, the preimage of the mapping $ \mathsf{M} $ is denoted as 
$\mathsf{M}^{-1}(\Pi \in {\cal P})$ which finds the set of all 
possible preference functions that converge to a Nash-stable 
partition $\Pi$. Thus, the Nash-stable core includes all those 
efficient allocation methods that build the following set:

\begin{equation}\label{NashStableCore}
{\cal N}\text{-}\mathrm{core} = \left\{ \Phi\in \mathbb{R}^{\kappa}  \left| \exists \Pi\in {\cal P} \right.; \mathsf{M}^{-1}\left(\Pi \right)  \ni {\Phi}\right\}.
\end{equation}

\subsection{Non-emptiness}
We define the set of constraints of a partition function. \textit{Efficiency} constraint dictates the following:
\begin{equation}
\mathcal{C}_{\mbox{Efficient}}(\Phi):=\left\{\sum_{i\in S}\phi_i^S = u(S), \forall S \subseteq N \right\}.
\end{equation} 
Given allocation method $\Phi$, we have the following constraints ensuring the Nash stability:
\begin{equation}
\mathcal{C}_{\mbox{Nash-stable}}(\Phi):=\left\{\phi_i^{S_{\mathsf{M}(\Phi)}(i)}\geq \phi_i^{T\cup i}, \forall T\in \mathsf{M}(\Phi)\cup \emptyset, \forall i \in N\right\},
\end{equation}
where $S_{\mathsf{M}(\Phi)}(i)$ is the unique set in $\mathsf{M}(\Phi)$ containing $i$. Then, the ${\cal N}$-core is non-empty, iif:
\begin{equation}
\exists \Phi^* :
\mathcal{C}_{\mbox{Efficient}}(\Phi^*) \mbox{ and } 
\mathcal{C}_{\mbox{Nash-stable}}(\Phi^*).
\end{equation}
The ${\cal N}$-core is defined as
\begin{equation}\label{NashStableCore}
{\cal N}\text{-}\mathrm{core} = \left\{ \Phi\in \mathbb{R}^{\kappa} : \mathcal{C}_{\mbox{Efficient}}(\Phi) \mbox{ and } \mathcal{C}_{\mbox{Nash-stable}}(\Phi)  \right\},
\end{equation}
which allows us to conclude:
\begin{theorem}
	The ${\cal N}$-core can be non-empty.
\end{theorem}
\begin{proof}
	The ${\cal N}$-core is non-empty if the following linear program is feasible:
	\begin{align}
	\min_{\Phi \in \mathbb{R}^{\kappa}} \Bigg\{ & \sum_{\forall S \subseteq N} 
	\sum_{\forall i \in S}\phi_i^S \mbox{ subject to }   \mathcal{C}_{\mbox{Efficient}}(\Phi) \mbox{ and } 
	\mathcal{C}_{\mbox{Nash-stable}}(\Phi) \Bigg\},
	\end{align}
	where we basically use optimization to find allocation methods which result in Nash-stable partitions.
\end{proof}
It may be hard to find non-emptiness of the 
Nash-stable core in the general case. Searching in an 
exhaustive manner over the whole partitions is NP-complete as the 
number of partitions grows according to the Bell number. Typically, 
with only $ 10 $ players, the number of partitions is as large as 
$115,975$.

We now analyze some specific cases in the following. 

\subsection{Grand Coalition Stability Conditions}
When the grand coalition $\Pi = \{N\}$ is targeted, the stability conditions are the following:
\begin{align}
\phi_i^N \geq u(i), \forall i\in N \mbox{ and }
\sum_{i\in N}\phi_i^{N} = u(N).
\end{align}
It is adequate to have $ u(N) \geq \sum_{i\in N} u(i) $ for the existence of allocation method guaranteeing the grand coalition. Those cooperative TU games that satisfy this condition are said to be \textit{essential}. 

\subsection{Symmetric Relative Gain}
We  propose to formulate a special case where the utility is shared among players with an equal relative gain. Let us denote the gain of player $i$ in coalition $S$ as $\phi_i^S = 
u(i) + \delta_i^S$ in which $\delta_i^S$ is called \textit{the 
	relative gain}. Note that for player $i$, one must have 
$\delta_i^i = 0$. The preference relation can be determined w.r.t. 
the relative gain, i.e., 
\begin{equation}
\delta_i^S \geq \delta_i^T \Leftrightarrow S \succeq_i T.
\end{equation}
The total allocated utilities in coalition $S$ is $\sum_{i\in 
	S}\phi_i^S = \sum_{i\in S} u(i) + \sum_{i\in S} \delta_i^S = u(S)$. 
Therefore, $\sum_{i\in S} \delta_i^S  = u(S) - \sum_{i\in S} u(i) 
=\Delta(S)$, where $\Delta(S)$ is the \textit{clustering profit} due 
to coalition $S$. The symmetric relative gain sharing relies 
on equally dividing the clustering profit in a coalition, i.e.
\begin{equation}\label{eq:equallyDivMarUtility}
\delta_i^S = \frac{\Delta(S)}{|S|}, \quad \forall i\in S.
\end{equation}
This choice means that each player in coalition $S$ has the same 
gain.

\begin{corollary}\label{cor:equivalentEvaluation}
	\textbf{Equivalent Evaluation}: Assume that $S\cap T \neq \emptyset$. 
	Due to  (\ref{eq:equallyDivMarUtility}), the following holds:
	\begin{equation}
	\frac{\Delta(S)}{|S|} \geq \frac{\Delta(T)}{|T|} \Leftrightarrow S\succeq_i T \quad \forall i\in S\cap T.
	\end{equation}
	It means that all players in $S\cap T$ prefer coalition $S$ to $T$ whenever the relative gain in $S$ is higher than $T$.
\end{corollary}
For this particular case, we have the following lemmas:
\begin{lemma}\label{lma:NashCoreTwoPlayersInCaseOfSymmetricRelativeGain}
	A game with two players $N=(1,2)$ always possesses a Nash-stable 
	partition in case of symmetric relative gain.
\end{lemma}
\begin{proof}
	see appendix 1.
\end{proof}

\begin{lemma}\label{lma:NashCoreThreePlayersInCaseOfSymmetricRelativeGain}
	A game with three players $N=(1,2,3)$ always possesses a Nash-stable 
	partition in case of symmetric relative gain.
\end{lemma}
\begin{proof}
	see appendix 2.
\end{proof}

Thus, we can conclude that symmetric relative gain always results in a Nash-stable partition when $n\leq 3$. However, this is not the case when $n>3$. We can find many counter examples that justify it such as the following one:
\begin{cexample}\label{cexample:NashCoreMoreThanThreePlayersInCaseOfSymmetricRelativeGain}
	Let the clustering profit for all possible $ S $ be as following: 
	\begin{align}
	&\Delta(1,2) = 0.86, \Delta(1,3) = 0.90, \Delta(1,4) = 0.87, 
	\Delta(2,3) = -1.22, \notag \\
	&\Delta(2,4) = -1.25, \Delta(3,4) = -1.21, \Delta(1,2,3) = 0.27, 
	\Delta(1,2,4) = 0.24, \notag \\
	&\Delta(1,3,4) = 0.28, \Delta(2,3,4) = -1.84, \Delta(1,2,3,4) = -0.35.
	\end{align}
	Let us now generate the preference profile according to these 
	clustering profit values. Note that we could eliminate those 
	clustering profit values which are negative since a player will 
	prefer to be alone instead of a negative relative gain. Further, 
	ranking the positive relative gains in a descending order results in 
	the following sequence:
	\begin{equation}
	\left[ \frac{\Delta(1,3)}{2}, \frac{\Delta(1,4)}{2}, \frac{\Delta(1,2)}{2}, \frac{\Delta(1,3,4)}{3}, \frac{\Delta(1,2,3)}{3}, \frac{\Delta(1,2,4)}{3} \right].
	\end{equation}
	According to the ranking sequence, we are able to generate the preference list of each player:
	\begin{align}
	& (1,3)\succ_1 (1,4)\succ_1 (1,2)\succ_1 (1,3,4)\succ_1 (1,2,3)\succ_1 (1,2,4) \succ_1 (1) \notag \\
	& (1,2)\succ_2 (1,2,3)\succ_2 (1,2,4) \succ_2 (2) \notag \\
	& (1,3)\succ_3 (1,3,4)\succ_3 (1,2,3)\succ_3 (3)\notag \\
	& (1,4)\succ_4 (1,3,4)\succ_4 (1,2,4) \succ_4 (4).
	\end{align}
	Note that this preference profile does not admit any Nash-stable 
	partition. Thus, we conclude that symmetric relative gain allocation 
	does not provide always a Nash-stable partition when $ n>3 $.
\end{cexample}

\subsection{Additively Separable and Symmetric Utility 
	Case}\label{sec:AdditivelySeparableandSymmetric}
We now focus on the case of separable and symmetric utilities. 
Consider equation (\ref{eq:additivelyseparable}) meaning that player 
$i$ gains $v_i(j)$ from player $j$ in any coalition. In case of 
symmetry, $v_i(j)=v_j(i)=v(i,j)$ such that $v(i,i)=0$. Further, we 
denote as $\phi_i^S = u(i) + \sum_{j\in S} v(i,j)$ the utility that 
player $i$ gains in coalition $S$. Then, the sum of allocated 
utilities in coalition $S$ is given by
\begin{equation}
\sum_{i\in S} \phi_i^S = \sum_{i,j\in S} v(i,j) + \sum_{i\in S}u(i) = u(S).
\end{equation}
Let us point out that $\sum_{i,j\in S} v(i,j) = 2\sum_{i,j\in S:j>i} v(i,j)$ (for example, $S=(1,2,3)$, $\sum_{i,j\in S} v(i,j) = 2[v(1,2) + v(1,3) + v(2,3)]$). Therefore, the following determines the existence of additively separable and symmetric preferences when the utility function $u$ is allocated to the players:
\begin{equation}
\sum_{i,j\in S:j>i} v(i,j) = \frac{1}{2}\left(u(S) - \sum_{i\in S} 
u(i)\right).
\end{equation}
Finding the values of $ v(i,j) $ satisfying these constraints are strongly restrictive and are rarely observed in real problems. Let it be illustrated for example $N=(1,2,3)$. The 
constraints imposed are:
\begin{align}
v(1,2) &= \frac{1}{2}[u(1,2)-u(1)-u(2)],\notag\\
v(1,3) &= \frac{1}{2}[u(1,3)-u(1)-u(3)],\notag\\
v(2,3) &= \frac{1}{2}[u(2,3)-u(2)-u(3)],\notag\\
v(1,2)+v(1,3)+v(2,3) &= \frac{1}{2}[u(1,2,3)-u(1)-u(2)-u(3)].
\end{align}
We have 3 variables and 4 constraints; thus, only special problem may 
fit with all the constraints. A more general approach allows to state 
the following theorem:
\begin{theorem}
	The ${\cal N}$-core of the additively separable and symmetric 
	utility hedonic game is non-empty if there exist balanced weights of 
	the dual problem (balancedness conditions):
	\begin{align}
	\exists \left\{w_S, \forall S\in 2^{N}\right\}:\\
	&\sum_{S\in 2^N}{w}_S \Delta(S) \leq \Delta(N), \notag \\
	&\sum_{S\in 2^N: {\cal V}(k)\in S} {w}_S = 1, \quad \forall k\in {\cal I},\notag \\
	&w_S \in [-\infty,\infty] \quad \forall S\in 2^N.
	\end{align}
\end{theorem}
\begin{proof}
	According to Bondareva-Shapley theorem 
	\cite{bondarevashapleytheorem1, bondarevashapleytheorem2} when the 
	gains of players are allocated according to the additively separable 
	and symmetric way, we have
	\begin{itemize}
		\item ${\cal V}$ the all possible bipartite coalitions such that ${\cal V}:= \{(i,j)\in 2^N: j>i\}$. Note that $|{\cal V}| = n(n-1)/2$.
		\item ${\cal I}$ the index set of all possible bipartite coalitions. So, ${\cal V}(k\in {\cal I})$ is the $k$th bipartite coalition.
		\item ${\bf v} = (v(i,j))_{(i,j)\in {\cal V}} \in \mathbb{R}^{|{\cal V}|}$ which is the vector demonstration of all $v(i,j)$.
		\item ${\bf 1} = (1,\ldots,1)\in \mathbb{R}^{|{\cal V}|}$.
		\item ${\bf b} = (b_S)_{S\in 2^N}\in \mathbb{R}^{2^n}$ such that $b_S = \frac{1}{2}\Delta(S)$ where $\Delta(S) = u(S) - \sum_{i\in S} u(i)$.
		\item ${\bf A} = (a_{S,k})_{S\in 2^N,k\in {\cal I}} \in \mathbb{R}^{2^n\times |{\cal V}|}$ is a matrix such that $a_{S,k} = \mathbb{1}\{S:{\cal V}(k)\in S\}$.
	\end{itemize}
	By using these definitions, the ${\cal N}$-core is non-empty whenever the following linear program is feasible
	\begin{align}
	({L})\quad &\min {\bf 1v} \quad \text{subject to} \notag\\
	&{\bf Av} = {\bf b}, {\bf v} \in [-\infty,\infty].
	\end{align}
	The linear program that is dual to $(L)$ is given by
	\begin{align}
	({\hat{L}})\quad &\max {\bf wb} \quad \text{subject to} \notag\\
	&{\bf wA} = {\bf 1}, {\bf w} \in [-\infty,\infty],
	\end{align}
	where ${\bf w} = (w_S)_{S\in 2^N}\in \mathbb{R}^{2^n}$ denote the vector of dual variables. Let ${\bf A}^{k}$ denote the $k$th column of ${\bf A}$. Then ${\bf w A}$ implies
	\begin{equation}
	{\bf w}{\bf A}^{k} = \sum_{S\in 2^N} w_S a_{S,k} = \sum_{S\in 2^N: {\cal V}(k)\in S} w_S = 1, \quad \forall k\in {\cal I}.
	\end{equation}
	This result means that the feasible solutions of $(\hat{L})$ exactly correspond to the vectors containing balancing weights for balanced families. More precisely, when ${\bf w}$ is feasible in $(\hat{L})$, ${\cal B}_{\bf w}$ is a balanced family with\textit{ balancing weights} $(w_S)_{S\in {\cal B}_{\bf w}}$.
	
	According to the \textit{weak duality theorem}, the objective function value of the primal $({L})$ at any feasible solution is always greater than or equal to the objective function value of the dual $(\hat{L})$ at any feasible solution, i.e. ${\bf 1}{\bf v} \geq {\bf w}{\bf b}$ which implies
	\begin{equation}
	{\bf 1}{\bf v} = \sum_{k\in {\cal I}}{v}_k = \sum_{\forall S\in 
		2^N}\sum_{i,j\in S: j>i} {v}(i,j) = \frac{1}{2}\Delta(N)
	\end{equation}
	and
	\begin{equation}
	{\bf w}{\bf b} = \frac{1}{2}\sum_{S\in 2^N}{w}_S \Delta(S) \leq \frac{1}{2}\Delta(N).
	\end{equation}
	
	Combining these results, we have the following balancedness conditions of $u$:
	\begin{align}
	&\sum_{S\in 2^N}{w}_S \Delta(S) \leq \Delta(N), \notag \\
	&\sum_{S\in 2^N: {\cal V}(k)\in S} {w}_S = 1, \quad \forall k\in {\cal I},\notag \\
	&w_S \in [-\infty,\infty] \quad \forall S\in 2^N.
	\end{align}
\end{proof}
However, these conditions are very restrictive. For a given set of players $n$, the number of variables is strictly equal to  $\binom{2}{n}=n(n-1)/2$ while the number of constraints is equal to the number of sets, i.e. $2^n-1$. For $n=10$, we have $45$ variables and as much as $512$ constraints.

\subsection{Relaxed Efficiency}
Considering the former result, we look at additively separable and symmetric utility case, by relaxing the efficiency constraint. Clearly, we relax the constraint of having the sum of allocated utilities in a coalition to be strictly equal to the utility of the coalition, i.e. 
$\sum_{i\in S}\phi_i^S \leq u(S)$ which leads to
\begin{equation}\label{eq:relaxedafficiency}
\sum_{i,j\in S:j>i} v(i,j) \leq \frac{1}{2}\Delta(S), \quad \forall S\in 2^N.
\end{equation}
The motivation behind relaxed efficiency can be the following: in case of the individual deviations, we do not favor the group interest; therefore, we can relax the efficiency condition (thus, we call it relaxed efficiency). It may be even interpreted as penalizing some coalitions to form. Thus, the following theorem may be stated:
\begin{theorem}
	The ${\cal N}$-core is always non-empty in case of relaxed efficiency.
\end{theorem}
\begin{proof}
	Finding the values of $ v(i,j) $ in eq. (\ref{eq:relaxedafficiency}) satisfying the relaxed efficiency condition can be done straightforward. However, we propose to formulate as an optimization problem for finding the values of $ v(i,j) $ so that $ \sum_{(i,j)\in {\cal V}} v(i,j) $ is maximized.
	
	A feasible solution of the following linear program guarantees the non-emptiness of the ${\cal N}$-core:
	\begin{align}\label{eq:nashStabelRelaxedEfficiency}
	&\max \sum_{(i,j)\in {\cal V}} v(i,j) \text{ subject to}\notag\\
	&\sum_{i,j\in S: j>i} v(i,j) \leq \frac{1}{2}\Delta(S), \forall S\in 2^N,
	\end{align}
	which is equivalent to
	\begin{align}
	({ LRE})\quad &\max {\bf 1v} \quad \text{subject to} \notag\\
	&{\bf Av} \leq {\bf b}, {\bf v} \in [-\infty,\infty].
	\end{align}
	Note that $({\rm LRE})$ is always feasible since
	\begin{itemize}
		\item there are no any inconsistent constraints, i.e. there are no at least two rows in ${\bf A}$ that are equivalent,
		\item the polytope is bounded in the direction of the gradient of the objective function ${\bf 1v}$.
	\end{itemize}
	
	For example, let $ N = (1,2,3,4) $. Then,
	\begin{align}
	&\max \left\{v(1,2)+v(1,3)+v(1,4)+v(2,3)+v(2,4)+v(3,4)\right\} \notag\\
	& \text{subject to}\notag\\
	& v(1,2)\leq \frac{\Delta(1,2)}{2}, \notag \\
	& v(1,3)\leq \frac{\Delta(1,3)}{2}, \notag \\
	& v(1,4)\leq \frac{\Delta(1,4)}{2}, \notag \\
	& v(2,3)\leq \frac{\Delta(2,3)}{2}, \notag \\
	& v(2,4)\leq \frac{\Delta(2,4)}{2}, \notag \\
	& v(3,4)\leq \frac{\Delta(3,4)}{2}, \notag \\
	& v(1,2)+v(1,3)+v(2,3)\leq \frac{\Delta(1,2,3)}{2}, \notag \\
	& v(1,2)+v(1,4)+v(2,4)\leq \frac{\Delta(1,2,4)}{2}, \notag \\
	& v(1,3)+v(1,4)+v(3,4)\leq \frac{\Delta(1,3,4)}{2}, \notag \\
	& v(1,2)+v(1,3)+v(2,3)\leq \frac{\Delta(2,3,4)}{2}, \notag \\
	& v(1,2)+v(1,3)+v(1,4)+v(2,3)+v(2,4)+v(3,4)\leq \frac{\Delta(1,2,3,4)}{2}  \notag \\
	& v(1,2)\in [-\infty,\infty], v(1,3)\in [-\infty,\infty], v(1,4)\in [-\infty,\infty], \notag \\
	& v(2,3)\in [-\infty,\infty], v(2,4)\in [-\infty,\infty], \notag \\
	& v(3,4)\in [-\infty,\infty]
	\end{align}
	Note that such a linear program has always a feasible solution. Moreover, for $ |N|>4 $, it is easily to conclude that the above linear program $ ({ LRE}) $ is always feasible.
\end{proof}


\section{Hedonic Coalition Formation as a Non-cooperative Game}
We can model the problem of finding a Nash-stable partition in hedonic coalition formation by formulating it as a non-cooperative game. We state the following:
\begin{theorem}
	A hedonic coalition formation game is equivalent to a non-cooperative 
	game.
\end{theorem}
\begin{proof}
	Let us denote as $\Sigma$ the set of strategies. We assume that the number of strategies is equal to the number of players, i.e. $ |\Sigma|=|N| $. This is sufficient to represent all possible choices. Indeed, the players that select the same strategy are interpreted as a coalition. For example, if every player chooses different strategies, then this corresponds to the coalition partition where every player is alone. 
	
	Consider the \textit{best-reply dynamics} where in a particular step $ s $, only one player chooses its best strategy. 
	A \textit{strategy tuple} in step $s$ is denoted as $\sigma^{s} = \{\sigma_1^{s}, \sigma_2^{s}, \ldots, \sigma_n^{s}\}$, where $\sigma_i^{s}$ is the strategy of player $i$ in step $s$. In every step, only one dimension is changed in $\sigma^{s}$. We further denote as $\Pi(\sigma^{s})$ the partition in step $s$. Define as $S_{i}^{(s)} = \{ j: \sigma_i^{s} = \sigma_j^{s}, \forall j\in N \}$ the set of players that share the same strategy with player $i$. Thus, note that $\cup_{i\in N} S_{i}^{(s)} = N$ for each step. The preference function of player $i$ is denoted as $\pi_i\left(\sigma^{s}\right)$ and verifies the following equivalence:
	\begin{equation}
		\pi_i\left(\sigma^{s}\right) \geq \pi_i\left(\sigma^{s-1}\right) \Leftrightarrow S_{i}^{(s)} \succeq_i S_{i}^{(s-1)}, 
	\end{equation}
	where player $i$ is the one that takes its turn in step $s$.
\end{proof}

\section{Conclusions}
In this work, we answered the following question: Is there any 
utility allocation method which could result in a Nash-stable 
partition? We proposed the definition of the Nash-stable core. We 
analyzed the cases in which the Nash-stable core is non-empty, and 
prove that in case of the relaxed efficiency condition there exists 
always a Nash-stable partition.

\section{Appendices}
\subsection{Appendix 1: proof of Lemma 
	\ref{lma:NashCoreTwoPlayersInCaseOfSymmetricRelativeGain}}
The proof is based on a simple enumeration of all possible partitions 
and corresponding conditions of Nash stability:
\begin{enumerate}
	\item $\Pi = \{(1),(2)\}$:
	\begin{align}
	& 0 \geq \delta_1^{12} \notag\\
	& 0 \geq \delta_2^{12} \notag\\
	& \delta_1^{12} + \delta_2^{12} = \Delta(1,2)
	\end{align}
	\item $\Pi = \{1,2\}$:
	\begin{align}
	& 0 \leq \delta_1^{12} \notag\\
	& 0 \leq \delta_2^{12} \notag\\
	& \delta_1^{12} + \delta_2^{12} = \Delta(1,2)
	\end{align}
\end{enumerate}
According to Corollary \ref{cor:equivalentEvaluation}, $\delta_1^{12} = \delta_2^{12} = \delta = \frac{\Delta(1,2)}{2}$. Thus, combining all constraint sets of all possible partitions, we have the following result constraint set: ${\cal C}_{\Pi} := \{0\leq\delta\}\cup \{0\geq\delta\} \Leftrightarrow \delta \in [-\infty,\infty]$. It means that for any value of $\Delta(1,2)$, symmetric relative gain always results in a Nash-stable partition for two players case.

\subsection{Appendix 2: proof of Lemma 
	\ref{lma:NashCoreThreePlayersInCaseOfSymmetricRelativeGain}}
Note that there are $5$ possible partitions in case of $N=(1,2,3)$. 
Thus, according to equally divided clustering profit, the following 
variables occur: $\delta_1^{12} = \delta_2^{12} = \delta_1$, 
$\delta_1^{13} = \delta_3^{13} = \delta_2$, $\delta_2^{23} = 
\delta_3^{23} = \delta_3$, $\delta_1^{123} = \delta_2^{123} = 
\delta_3^{123} = \delta_4$. Enumerating all possible partitions 
results in the following conditions:
\begin{enumerate}
	\item $\Pi = \{(1),(2),(3)\}$:
	\begin{equation}
	\delta_1\leq 0, \delta_2\leq 0, \delta_3\leq 0,
	\end{equation}
	\item $\Pi = \{(1,2),(3)\}$:
	\begin{equation}
	\delta_1\geq 0, \delta_1\geq \delta_2, \delta_1\geq\delta_3, \delta_4\leq 0,
	\end{equation}
	\item $\Pi = \{(1,3),(2)\}$:
	\begin{equation}
	\delta_2\geq 0, \delta_2\geq \delta_1, \delta_2\geq\delta_3, \delta_4\leq 0,
	\end{equation}
	\item $\Pi = \{(2,3),(1)\}$:
	\begin{equation}
	\delta_3\geq 0, \delta_3\geq \delta_1, \delta_3\geq\delta_2, \delta_4\leq 0,
	\end{equation}
	\item $\Pi = \{1,2,3\}$:
	\begin{equation}
	\delta_4\geq 0.
	\end{equation}
\end{enumerate}
Note that the constraint set ${\cal C}_{\Pi}$ covers all values in $\delta_1, \delta_2, \delta_3$ in case of $\delta_4 \geq 0$. Further, it also covers all values when $\delta_4 \leq 0$. We are able to draw it since there are three dimensions:
\begin{equation*}
\includegraphics[width=\linewidth]{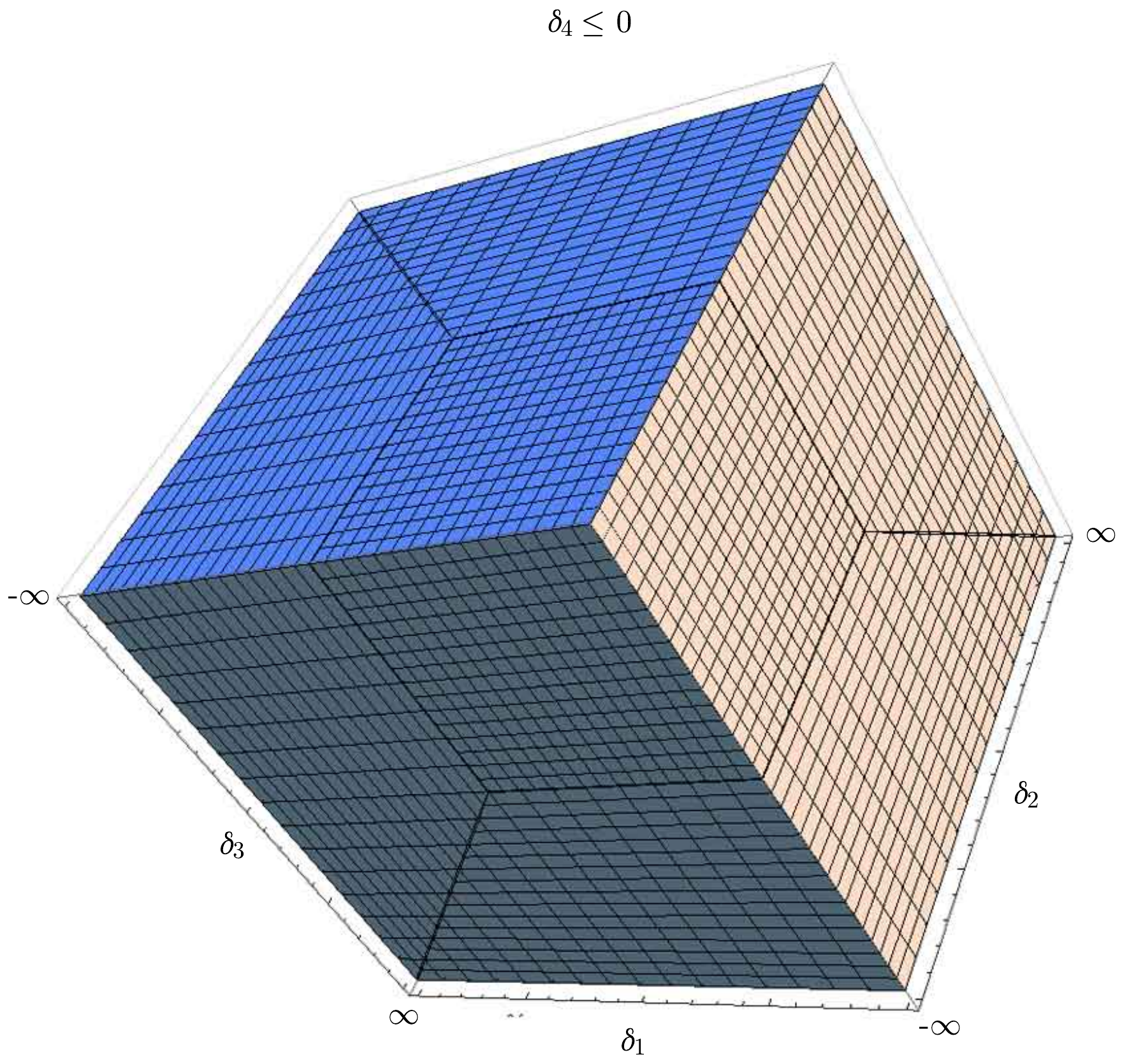}
\end{equation*}

\end{document}